# Direct identification of monolayer rhenium diselenide by an individual diffraction pattern

Zhen Fei[1], Bo Wang[2], Ching-Hwa Ho[3], Fang Lin[4], Jun Yuan[5,1], Ze Zhang[1] and Chuanhong Jin[1](✉)

[1] State Key Laboratory of Silicon Materials, School of Materials Science and Engineering, Zhejiang University, Hangzhou, 310027, China
[2] School of Physical Science and Technology, Lanzhou University, Lanzhou, Gansu 730000, China
[3] Graduate Institute of Applied Science and Technology, National Taiwan University of Science and Technology, Taipei 106, Taiwan
[4] College of Electronic Engineering, South China Agricultural University, Guangzhou, Guangdong 510642, China
[5] Department of Physics, University of York, Heslington, York YO10 5DD, UK

## ABSTRACT

In the current extensive studies of layered two-dimensional (2D) materials, compared to the hexagonal ones, like graphene, hBN and $MoS_2$, low symmetry 2D materials have shown great potential for applications in anisotropic devices. Rhenium diselenide ($ReSe_2$) has the bulk space group $P\bar{1}$ and belongs to triclinic crystal system with a deformed cadmium iodide type structure. Here we propose an electron diffraction based method to distinguish monolayer $ReSe_2$ membrane from multilayer $ReSe_2$, and its two different vertical orientations, our method could also be applicable to other low symmetry crystal systems, including both triclinic and monoclinic lattices, as long as their third unit-cell basis vectors are not perpendicular to their basal planes. Our experimental results are well explained by kinematical electron diffraction theory and corresponding simulations. The generalization of our method to other 2D materials, like graphene, is also discussed.

## 1 Introduction

Since the successful fabrication of atomically thin carbon film field effect transistor, there has been growing interest in investigating van der Waals layered materials among the research communities[1]. In particular, the family of transition metal dichalcogenides (TMDCs), when reduced to atomic thicknesses, shows great potential for future applications in semiconducting electronics and optoelectronics, and also opened up more opportunities to explore novel physical phenomena in condensed matter physics, such as valley physics[2, 3], superconductivity[4], charge density waves[5, 6], and large magnetoresistance[7].

Of the layered 2D materials, low symmetry 2D materials have emerged as anisotropic electronic and optoelectronic candidates. Compared to graphene and $MoS_2$ with the highly symmetric hexagonal structures, those 2D materials with reduced symmetry, ranging from black phosphorus[8-10] and $WTe_2$[11] to $ReS_2$ and $ReSe_2$, have internal anisotropic physical properties.

Address correspondence to Chuanhong Jin, chhjin@zju.edu.cn

Moreover, ReS$_2$ and ReSe$_2$ share similar crystal structures, they both have a distorted octahedral 1T structure and belong to triclinic crystal system, and Re atoms form zigzag chains in the basal plane, arising from the Peierls distortion. Therefore, ReSe$_2$ and ReS$_2$ flakes have both in-plane and out-of-plane anisotropy, and recently there have been numbers of applications based on their anisotropic properties [12-17].

When materials are thinned down to atomically thin membranes, thickness begins to play an important role in tailoring their properties, and vice versa, there are a variety of techniques to determine their thicknesses, ranging from image contrast of reflected light microscopy[18-24], second harmonic microscopy[25-28], intuitive atomic force microscopy and cross-sectional imaging to Raman and photoluminescence spectroscopy[29-32]. In the field of electron microscopy, there are miscellaneous methods to identify thickness as well, which include peak shift in plasmon spectroscopy[33, 34], direct high-resolution transmission electron microscopy (HRTEM) imaging combined with simulations[35] and linearity in annular dark field scanning transmission electron microscope (ADF-STEM) signals[36, 37]. Apart from the above methods, electron diffraction analysis [34, 38-43] has also been demonstrated as an efficient tool to determine thickness in 2D materials, since it could be conducted on any commercial uncorrected TEM conveniently with negligible beam damages on a large area of pristine crystalline samples. Furthermore, diffraction based methods usually involve a series of sample tilting or use the relative ratio of the chosen diffraction spots, and some researchers noticed the intensity mismatch in a pair of crystallographically equivalent diffraction spots (Friedel pair)[41, 44, 45], here we demonstrate a method of identifying monolayer ReSe$_2$ by the centrosymmetry with a single diffraction pattern.

## 2  Theoretical basis

To illustrate the essence of electron diffraction pattern evolution with thickness of atomically thin ReSe$_2$, an analytic kinematical diffraction theory approximation has been developed in this study, which is usually applicable to ultra-thin specimens and light-element atoms[38, 41]. In the classic description of electron diffraction processes, electrons are scattered by specimen's electrostatic potential, and according to the first Born approximation, amplitude of diffracted beam is proportional to the Fourier transform of the corresponding specimen's potential. Furthermore, without considering the effect of interatomic bonding, the total specimen's potential could be written as a superposition of all individual atoms' potentials and the Fourier transform of one atom's potential is its atomic scattering factor[46-48].

In the case of few-layer ReSe$_2$, we first take summations of atomic scattering factors over a unit cell, containing four rhenium and eight selenium atoms as shown in Fig. 1(a)&(b), and then combine all unit cells along the thickness direction, the two terms, as a whole, are defined as a basis sum term. Then we sum this term laterally over the entire basal plane, defined as the two-dimensional lattice summation term[38]. This is formulated as

$$f(\vec{K}) = \sum_{lattice} e^{-i2\pi \vec{K}\cdot \vec{R}_l} \sum_{basis} e^{-i2\pi \vec{K}\cdot \vec{R}_b} f_{atom(Re,Se)}(\vec{K}) \quad (1)$$

where $R_l$ and $R_b$ is the discrete lattice and basis vector of ReSe$_2$ crystal respectively, $\vec{K}$ is the difference between diffracted and incident electron wave vector, $f_{atom(Re,Se)}(\vec{K})$ is the atomic scattering factor for Re or Se atoms, and this formula could be further simplified as

$$f(\vec{K}) = s_{lattice}(\vec{K}) s_{stacking}(\vec{K}) F_{structure}(\vec{K}) \quad (2)$$

where the three terms represent summations over 2D lattice plane, vertical direction and one unit cell respectively. We will discuss in detail about the multiplication of the three terms. The first term is the result of in-plane periodicity of layered ReSe2 crystal, by taking square of the modulus, and hence,

$$|s_{lattice}(\vec{K})|^2 = \frac{sin^2(\frac{n_1 2\pi \vec{K} \cdot \vec{a}}{2}) sin^2(\frac{n_2 2\pi \vec{K} \cdot \vec{b}}{2})}{sin^2(\frac{2\pi \vec{K} \cdot \vec{a}}{2}) sin^2(\frac{2\pi \vec{K} \cdot \vec{b}}{2})} \quad (3)$$

where $\vec{a}$ and $\vec{b}$ are in-plane lattice vectors indicated in Fig. 1(a)&(b), $n_1$ and $n_2$ are the number of unit cells along their corresponding directions. This equation could be treated as a Dirac delta function, $|s_{lattice}(\vec{K})|^2 \sim n_1 n_2 \delta(\alpha)\delta(\beta)$ as long as $n_1$ and $n_2$ are large, in other words, the illuminated area is large enough. $\alpha$ and $\beta$ represent deviations from the nearest reciprocal lattice points along basal plane directions. Therefore, this term is none-zero only for $\vec{K} = \vec{G} + \vec{k_\perp}$, where $\vec{G} = h\vec{a}^* + k\vec{b}^* + l\vec{c}^*$, representing the reciprocal lattice points, and $\vec{k_\perp}$ is the perpendicular component of the vector describing deviation from the reciprocal lattice point. Therefore, none-zero values are often interpreted as reciprocal rel-rods. The second term is determined by the stacking of unit cells in the out-of-plane direction, expressed as

$$|s_{stacking}(\vec{K})|^2 = \frac{sin^2(\frac{2\pi N\vec{K} \cdot \vec{c}}{2})}{sin^2(\frac{2\pi \vec{K} \cdot \vec{c}}{2})} \quad (4)$$

which is similar to the first term, but N is much smaller since it represents the layer number of the corresponding atomically thin layered ReSe2 crystal. Therefore, the value of the second term, also defined as the shape factor in conventional bulk crystal, will oscillate along the z direction, with the periodicity of $c^*$, the length of $\vec{c}^*$, but remain constant for monolayer. All direction specifications in our discussions, involving xyz-Cartesian coordinates, are consistent with the illustration in Fig. 1(a)&(b). The last term is named as the structure factor, by simply summing up all atomic factors in one unit cell, which could be calculated numerically according to the inversion symmetry inside the unit cell, then

$$|F_{structure}(\vec{K})|^2 = 4[f_{Re}(\vec{K})\sum_{i=1}^{2} cos2\pi(h\Delta a_i + k\Delta b_i + l\Delta c_i) + \\ f_{Se}(\vec{K})\sum_{j=1}^{4} cos2\pi(h\Delta a_j + k\Delta b_j + l\Delta c_j)]^2 \quad (5)$$

where $\Delta a_i$, $\Delta b_i$ and $\Delta c_i$ represents Re atom's relative coordinate deviations from the inversion center of unit cell in the $\vec{a}$, $\vec{b}$ and $\vec{c}$ direction respectively, therefore, in an unit cell, there are two and four crystallographically inequivalent atoms for Re and Se respectively, and their atomic scattering factors are well tabulated[46].

In brief, the first term restricts none-zero values to the rel-rods in reciprocal space, which are located at the reciprocal lattice points and perpendicular to the basal plane, then the shape factors modulate the intensity along the thickness direction, with a characteristic length of $c^*$, finally the structure factors will tune the intensity much more slowly in z direction, and at different (h, k) points in the basal plane. Here, we are mainly focusing on the shape factor, an indicator of the sample thickness. In Fig. 1(c), the shape factors of a bi-layer ReSe2 are visualized in the reciprocal space. As we can see, the plane formed by vector $\vec{a}^*$ and $\vec{b}^*$ is inclined in the reciprocal space, as the third unit cell basis vector in direct space is not perpendicular to the basal plane and those inclined planes will intersect with basal plane along a series of parallel lines. More specifically, the distances of the nearest reciprocal lattice points with basal plane will vary periodically, as illustrated in Fig. 1(d).

As we have discussed in the kinematical diffraction theory, electron diffraction pattern could be interpreted as the intersections between Ewald sphere and reciprocal rel-rods, also known as the multiplication of the shape factors and the structure factors. Figure. 2(a) shows a pair of crystallographically equivalent shape factors and their intersections with Ewald sphere. It is noted that the two intersections are not symmetrical, and the distances between the intersections and centroids of the shape factor (reciprocal lattice points) are written as $\Delta_1$ and $\Delta_2$ respectively, which determine their corresponding intensities. In addition, the distance between Ewald sphere and basal plane is denoted as $\Delta_c$, and the distance between the centroid of the shape factor and basal plane is $\Delta_s$, therefore, $\Delta_1 = |\Delta_s - \Delta_c|$ and $\Delta_2 = |\Delta_s + \Delta_c|$. Such a mismatch could be attributed to the curvature effect of Ewald sphere and the inclined nature of $\vec{a^*}$ and $\vec{b^*}$, the reciprocal basis vectors in the triclinic crystal lattice. In other words, when the incident beam is perpendicular to the basal plane, the ReSe$_2$ crystal do not sit in the zone axis direction, and thus the prerequisite of well-known Friedel law could not be satisfied, hence, there is a mismatch in the intensity for a pair of crystallographically equivalent spots except for monolayer. As illustrated in Fig. 2(b), $\Delta_s$ is determined by the vertical position of the reciprocal lattice point, as we have shown in Fig. 1(c)&(d), $\Delta_c$ is determined by the curvature of Ewald sphere, which will increase when it deviates away from the origin. Moreover, $\Delta_c$ is usually small compared to $c^*$, which could be regarded as a small increment or decrement along the shape factor, and thus, such a mismatch is approximately proportional to the derivative of the intensity of shape factor with respect to z-direction, as shown in Fig. 2(b). Moreover, such a mismatch could also be interpreted as a result of phase shift in the intensity vs tilt-angle curve. Figure. 2 (c)-(f) display how the simulated intensities of ReSe$_2$ with different thicknesses vary with tilt-angle, we have chosen several pairs of diffraction spots to explore the 3D information of the reciprocal space. It is now clear that compared to multilayer, the selected diffraction spot intensities of monolayer ReSe$_2$ are insensitive to tilt-angle and the corresponding Friedel pairs have negligible intensity mismatch. More details about the effects of tilting monolayer ReSe$_2$ are discussed in Fig. S-1 in Electronic Supplementary Material (ESM). Furthermore, notice the phase shift $\Delta_s$ will cause the two curves to separate from each other, and thus the intensity mismatch. Such a mismatch is dependent on the origin position of the curve, in other words, the intersection of Ewald sphere and shape factor. For example, reciprocal lattice point pair of $(\bar{2}\bar{3}1)\&(23\bar{1})$ are far from basal plane, and the corresponding origin positions of them are shifted into a fast oscillation zone, as illustrated in Fig. 2(e). As a result, such a mismatch at origin increases when layer number goes from monolayer to bilayer, and drops down at 3L, then rises again for 4L. It is worth to point out that with the increase in distance of diffraction spot from the tilt-axis, the sensitivity to tilt-angle increases, since the corresponding tilt-arm will increase, as indicated in Fig. 2(f), the modulation period is larger than that of (e). Apart from choosing different Friedel pairs, increasing Ewald sphere curvature could also broaden intensity mismatch, see more discussions in Fig. S-2 in ESM.

Through our discussions on the theory and simulations, we demonstrates that, by figuring out the intensity mismatch between a Friedel pair, we could retrieve part of the 3D information in the reciprocal space without tilting our ReSe$_2$ crystal, therefore, our method could be used to unambiguously identify monolayer ReSe$_2$ crystalline with one single diffraction pattern.

## 3   Results and discussion
Atomically resolved ADF-STEM image was employed to confirm monolayer ReSe$_2$ sample, as a reference to demonstrate our following diffraction based method. In Fig. 3(a), atoms of rhenium and selenide could be clearly resolved, and brighter atoms represent rhenium atoms since heavier atoms scatter electrons more strongly. In the left side of this figure, diamond-shaped-chains, a fingerprint of monolayer ReSe$_2$, is distinctively distinguishable, as illustrated by the superimposed atomic model (inset is the corresponding low magnification ADF-STEM image). Once the monolayer region was confirmed, different thicknesses could be identified according to the step-like intensity profiles, as shown in Fig. 3(b)[36, 37]. Therefore, this method was employed to cross check the thickness of few-layer ReSe$_2$ membranes.

ReSe$_2$ belongs to triclinic crystal system, which has the lowest symmetry of all, and lacks the symmetry of C$_2$. In other words, flipping ReSe$_2$ membrane over is not crystallographically equivalent under the same experimental configuration, and we have to determine its vertical orientation at first. Figure. 4(a)&(b) display simulated diffraction patterns for upside and downside oriented monolayers ReSe$_2$ respectively, which are obtained by flipping around the y-axis. As we discussed earlier, the shape factors of monolayer remain constant along z-direction, and the resultant diffraction pattern should flip around the same axis as the crystal does, but the corresponding multi-layer ReSe$_2$ behave differently, they flip around the horizontal axis, the x-axis, more details are discussed in Fig. S-3 in ESM. Experimental diffraction patterns for the corresponding two vertical orientations are shown in Fig. 4(c)&(d), and they are consistent with simulated ones. Next, for the concern of signal noise ratio, lower order diffraction spots suffer from the background noise of central beam, hence we select Friedel pairs away from central beam and those with high brightness. Figure 4(e) shows intensity mismatch of the chosen Friedel pairs in different layers from experimental diffraction patterns, intensity mismatch is defined as $|I_1 - I_2| / (I_1 + I_2)$, where $I_1$ and $I_2$ are the integrated intensity of a Friedel pair (background subtracted). We find that intensity mismatch is negligible in monolayer ReSe$_2$, and becomes significant in multilayers, but the trends are not always monotonically increasing and the error bars increase dramatically since multilayers are more sensitive to small tilt angles (see more discussions on the effects of different tilt-axis in Fig. S-4 in ESM). In particular, the pair of $(\bar{2}\bar{3}1)$&$(23\bar{1})$, whose intensity mismatch experienced ups and downs from monolayer to four-layer, behaves exactly as expected from the previous discussions in Fig. 2(e). For full comparisons of simulated and experimental diffraction patterns for different thicknesses and extended discussions, see Fig. S-5 in ESM. Finally, our method is not limited to ReSe$_2$, from the derivation of the theory, we learned that the layered crystals of triclinic and monoclinic systems will follow the same rule, since their third unit-cell basis vectors are also not perpendicular to the basal plane. Moreover, for high symmetry layered crystals, like graphene, our method will still work if they are in the AB-stacking configuration, and the generalization to AA-stacking requires tilting sample off the zone axis, details are discussed in Fig. S-6 in ESM.

In conclusion, by measuring the intensity difference between a pair of crystallographically equivalent diffraction spots, we could distinguish monolayer from multi-layer ReSe$_2$ membranes, and identify its vertical orientations.

*Generally, inspired by the method of tilt-series diffraction analysis to determine thickness of atomically thin crystals, we took advantage of the intrinsic low symmetry property of ReSe$_2$ crystal structure, and acquired an individual diffraction pattern containing the information of intensity mismatch of a pair of crystallographically equivalent spots, to identify monolayer ReSe$_2$. Furthermore, by visually comparing the diffraction patterns of monolayers ReSe$_2$, we confirmed its two vertical orientations, which*

*could be exploited for anisotropic applications. More importantly, this method, in principle, could be generalized to other 2D materials with triclinic or monoclinic lattice structures, and even graphene-like high symmetry 2D materials.*

## 4  Conclusions

We conclude that by quantitatively analyzing electron diffraction pattern of few-layer $ReSe_2$ membranes, it is possible to identify monolayer $ReSe_2$ and its vertical orientations without taking tilt-series diffraction patterns, as an alternative to the previous reported thickness determination method by electron diffraction. In our study, we have distinguished monolayer $ReSe_2$ from multilayer according to the intensity mismatches of Friedel pairs, and determined its vertical orientations through the low symmetric diffraction patterns. This analytical technique is based on the low symmetry of $ReSe_2$ crystal structure and the curvature of Ewald sphere, finally it could be applicable to crystals of both triclinic system and monoclinic system, as long as their third unit-cell basis vectors are not perpendicular to the basal plane, and generalized to even those high symmetry 2D materials.

## Methods

Layered crystals of $ReSe_2$ were grown by chemical-vapor transport method using $I_2$ as the transport agent. The method consisted of two steps: First, prior to the crystal growth the powdered compound of the rhenium diselenide was prepared from the elements (Re: 99.95% and Se: 99.999%) by reaction at 1030 °C for 10 days in evacuated quartz ampoules. About 10 g of the elements were introduced into a quartz ampoule (19mm OD, 14mm ID, 15 cm length), which was then evacuated to about $10^{-6}$ torr and sealed. For crystal growth, the chemical transport was achieved by an appropriate amount (~10 g) of synthesized compound together with the transport agent ($I_2$ about 10 mg/cm$^3$) placed in a quartz ampoule (22mm OD, 17mm ID, and 20 cm in length), which was then cooled with liquid nitrogen, evacuated to $10^{-6}$ torr and sealed. The quartz tube was then placed in a horizontal three-zone furnace and the charge pre-reacted for 24 h at 800 °C while the temperature of the growth zone was set at 1000 °C to prevent the transport of the product. The furnace was slowly set to give a constant temperature of 1000 °C across the reaction tube, and then programmed over 24 h to produce the temperature gradient at which single crystal growth takes place. Best results were obtained with temperature setting of about 1060→1010 °C in a temperature gradient of about -2.5 °C/cm and growth time of about 20 days.

To prepare free-standing few-layer $ReSe_2$ membranes, polydimethylsiloxane (PDMS) was used for micromechanical exfoliation of bulk high quality $ReSe_2$ crystals, and then thin flakes were deposited on top of an oxidized silicon wafer. Later, few-layer membranes were identified by optical microscope and the desired region was carefully aligned to holes of a lacey carbon TEM grid. Then, acetone was used to join carbon film and sample through evaporation of acetone in air, subsequently, sodium hydroxide (NaOH) solution was used to etch away the underneath $SiO_2$ layer. Finally, the TEM grid with sample on it was immersed in water to dissolve residual NaOH solute, and subsequently in acetone to prevent support film from rupturing, after dried in air, our sample was ready for further characterizations.

Diffraction patterns of specimen were recorded on a FEI Tecnai G-F20 microscope operated at 200 kV. The uniform lateral dimension of few-layer membrane is on the scale of few microns, therefore an illuminated area of 200 nm in diameter was chosen by selected area aperture, to ensure electrons are from the region of interest. To obtain diffractions of the specimen with opposed vertical directions, the TEM grid was reloaded and flipped over outside the microscope. For quantitative diffraction intensity analysis, the exposure time was limited in the range where the response function of CCD detector remain linear.

The ADF-STEM images were recorded on a probe-corrected Titan ChemiSTEM at an acceleration voltage of 200 kV. After acquiring the atomically resolved images, we further processed images by a technique of superimposing a stack of drift corrected images to enhance signal to noise ratio for better visibility, hence, monolayer and multi-layer $ReSe_2$ membranes can be unambiguously distinguished, more specifically, multilayer $ReSe_2$ are stacked with a lateral shift in basal plane, and the in-plane motif of diamond-shaped-chains for monolayer will be blurred as layer number increases. In addition, intensity profiles extracted from low-magnification ADF-STEM image could be employed to determine the corresponding relative thickness, since the number of large-angle scattered electrons is approximately proportional to the number of atoms involved.

Electron diffraction simulations were completed with the open source multislice simulation software package QSTEM[49]. At first, large enough atomic structures with different orientations (varying tilt-angle and tilt-axis) were exported from the software of CrystalMaker, and they were then feed to QSTEM to generate corresponding diffraction patterns, the above processes were automated with a home-made Python script. Under the settings of QSTEM, a box with dimensions of 300Å ×300Å ×100Å (150Å ×150Å ×100Å for large angle tilting) was used and the slice thickness was 4.7Å, with the tilt-angle ranging from -5° to 5° in steps of 1° (from -30° to 30° in steps of 2° for large angle tilt), with the in-plane rotation of tilt-axis ranging from 15° to 360° in steps of 15°. In addition, a 1k×1k pixel$^2$ array was adopted (500×500 for large angle tilting) to keep a resolution of 0.3Å, and the acceleration voltage was set to 200 kV (60 kV for low voltage simulations).

# Acknowledgements

This work is financially supported by the National Basic Research Program of China (Grant No. 2014CB932500 and No. 2015CB921004), the National Natural Science Foundation of China (Grant No. 51472215, No. 51222202, No. 61571197 and No. 61172011), the 111 project (No. B16042) and the Ministry of Science and Technology, Taiwan (MOST 104-2112-M-011-002-MY3). The authors would like to thank Prof. Christoph Koch from Humbold University of Berlin for the fruitful discussions on multislice simulations. J.Y. acknowledges the EPSRC (UK) funding EP/G070326 and EP/J022098 and supports from Pao Yu-Kong International Foundation for a Chair Professorship in ZJU. This work made use of the resources of the Center of Electron Microscopy of Zhejiang University.

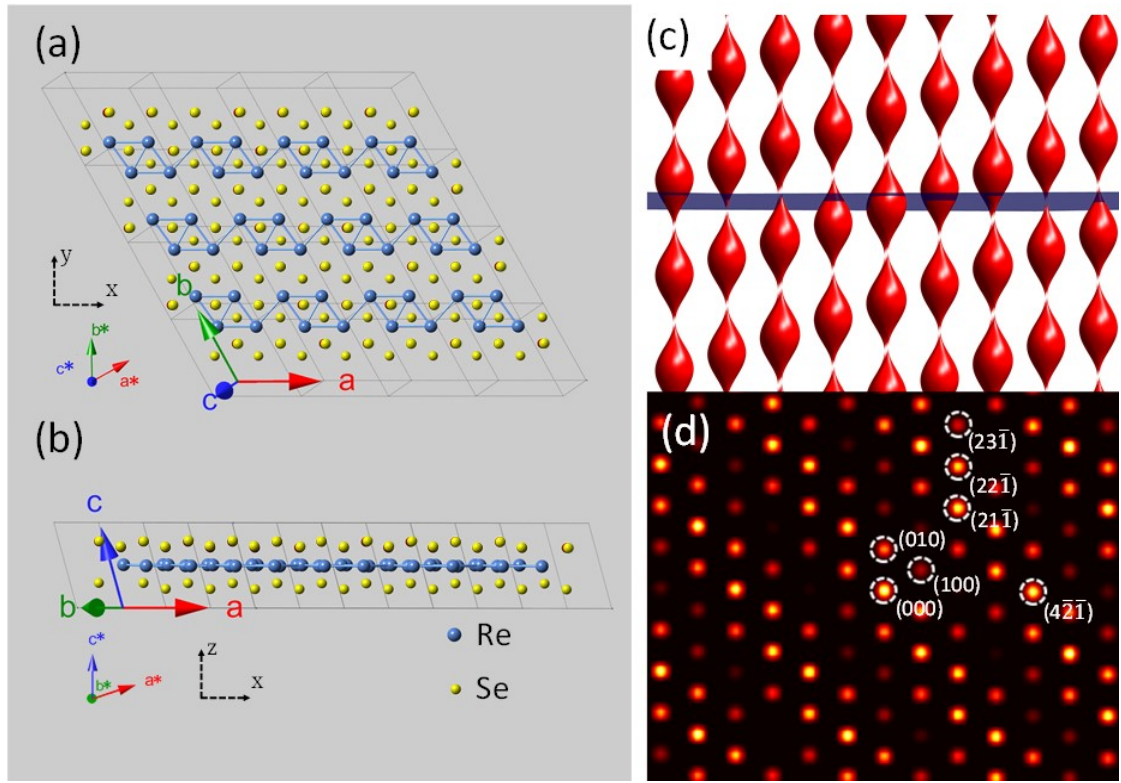

**Figure 1** Structure of ReSe$_2$ in direct and reciprocal space: (a) Atomic structure viewed from z-direction in Cartesian coordinates, unit cell basis vectors are coded with different colors, and reciprocal unit cell basis vectors are indicated by superscripts stars. (b) Atomic structure viewed from y-direction. (c) Visualization of shape factors in reciprocal space for the bi-layer ReSe$_2$ crystal. (d) Distances of the nearest reciprocal lattice points with basal plane, the closer the point is from basal plane, the brighter the corresponding spot is.

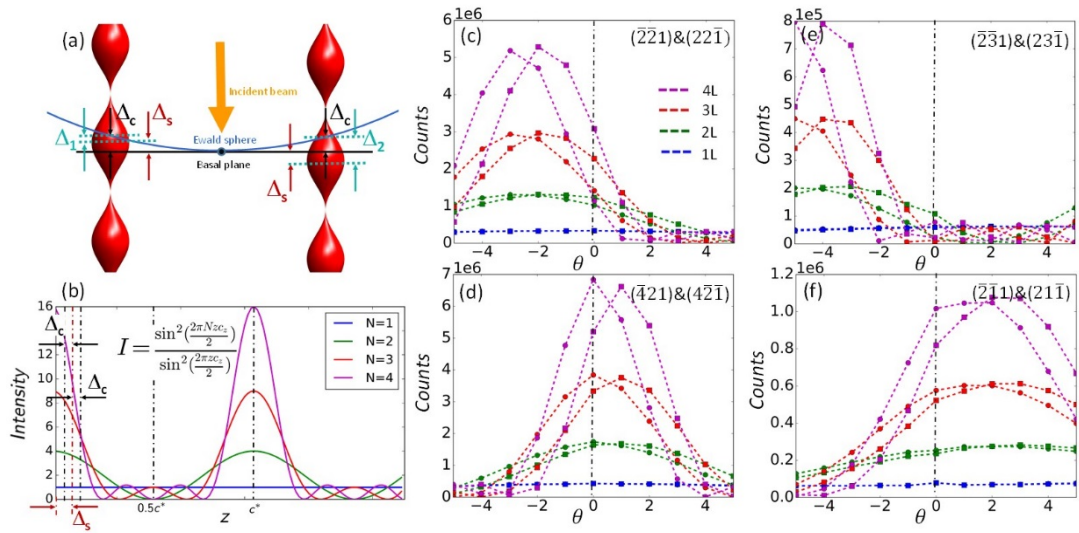

**Figure 2** (a) Schematic description of the asymmetrical intersections between a pair of crystallographically equivalent shape factors and Ewald sphere. (b) Intensity modulatoin of shape factor in z-direction for different thicknesses, where $c_z$ is the projection of $\vec{c}$ onto z-axis. (c)-(f) Simulated intensity modulations with tilt-angle of selected Friedel pairs for different thicknesses, simulated diffraction patterns of (c), (e)&(f) are tilted around x-axis, while (d) is tilted around y-axis.

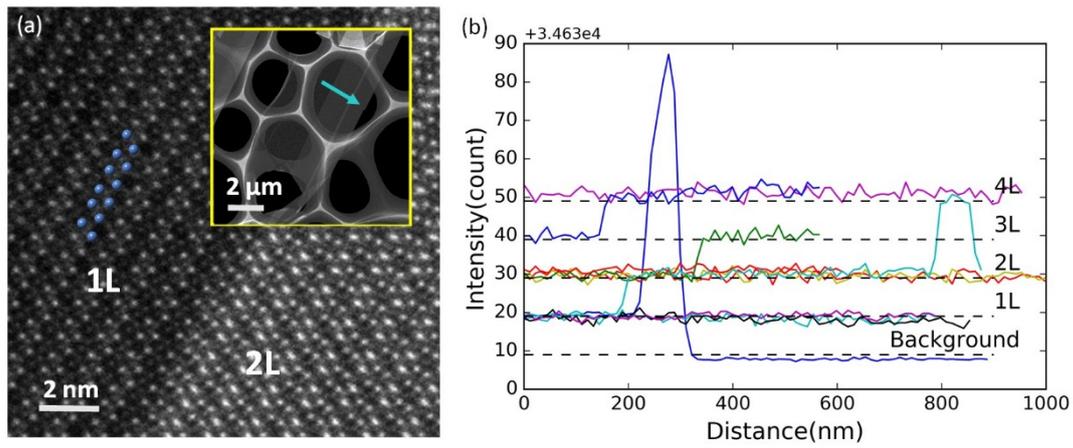

**Figure 3** (a) Atomically resolved image of monolayer and bilayer ReSe$_2$, inset is the corresponding low-magnification ADF-STEM image. (b) Step-like intensity profiles for different thicknesses, the greenish line is extracted along the greenish arrow in the inset of (a).

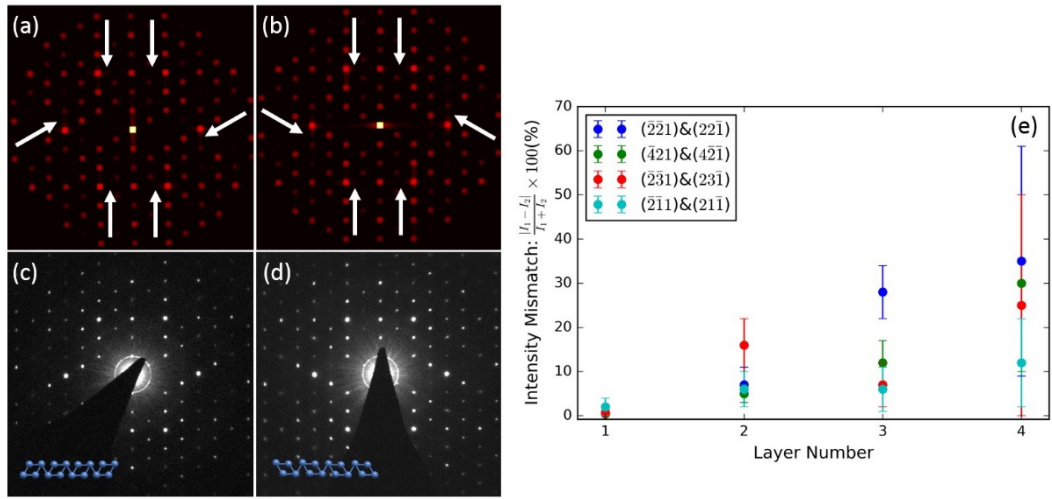

**Figure 4** (a)&(b) Simulated electron diffraction patterns of monolayer ReSe$_2$ for upside and downside orientations respectively, white arrows indicate how intensity varies for better visual comparisons. (c)&(d) Experimental electron diffraction patterns of monolayer ReSe$_2$ for upside and downside orientations respectively. (e) Intensity mismatch of selected Friedel pairs from experimental electron diffraction patterns.

# Electronic Supplementary Material

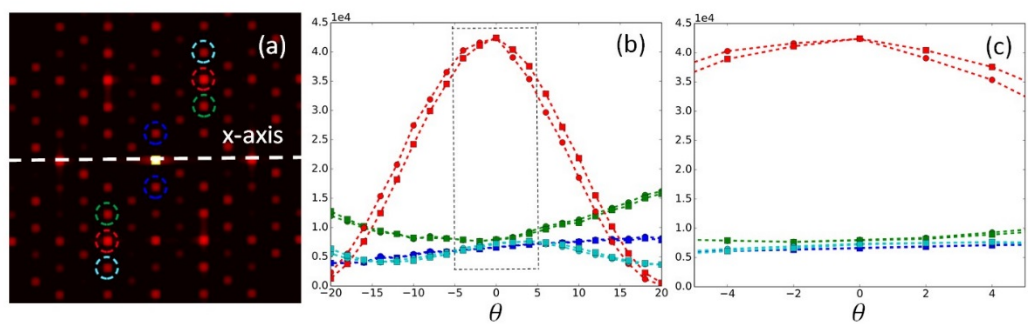

**Figure S1** Large-angle tilt diffraction pattern series were simulated for monolayer ReSe$_2$ to explore its structure factors since monolayer ReSe$_2$ is a sandwiched structure with three layers closely packed, instead of an ideal monolayer. (a) one of the simulated diffraction patterns, and the tilt axis is x-axis, with selected spots coded with different colors, (b) intensities of selected diffraction spots in (a) vary with tilt-angle, up to $\pm 20°$, zoom-in area indicated by the dashed black box is displayed in (c). As compared to shape factors, structure factor variation is much slower for monolayer ReSe$_2$ and there is no intensity mismatch at zero tilt-angle.

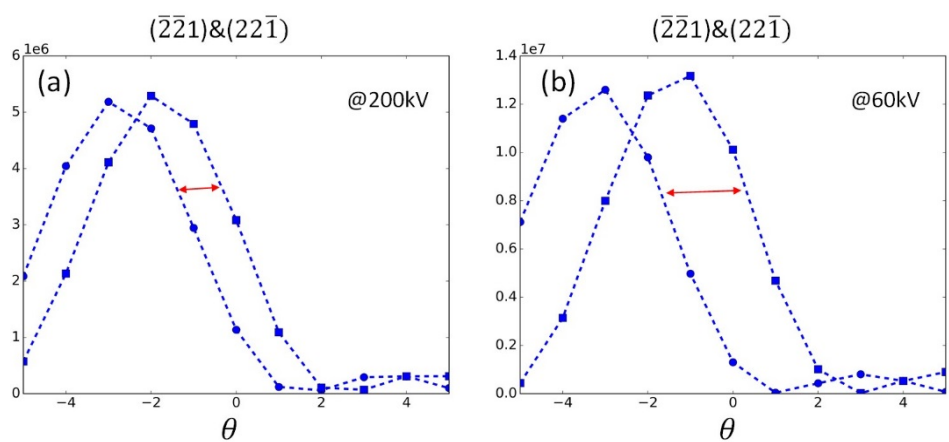

**Figure S2** (a)&(b) Intensity variations, of the selected Friedel pair $(\bar{2}\bar{2}1)\&(22\bar{1})$ in 4L ReSe$_2$, with tilt-angle, which are from simulated diffraction patterns at accelerating voltages of 200 kV and 60 kV respectively. Intensity mismatch could be broadened by increasing the curvature of Ewald sphere, which is to decrease the accelerating voltage.

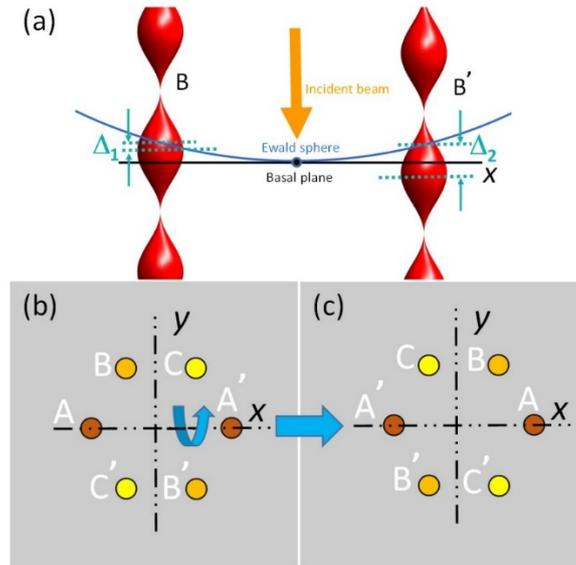

**Figure S3** Illustration of how the intensity of electron diffraction pattern in multilayer ReSe$_2$ change when crystal was flipped upside down. When the Friedel pair in (a) was flipped around x-axis in (b), not only their positions flipped over to the opposite side of x-axis, but their intensities exchanged with each other, which was shown in (c), the diffraction pattern seemed to have flipped around y-axis (perpendicular to x-axis) if we have only considered the intensity of each spot.

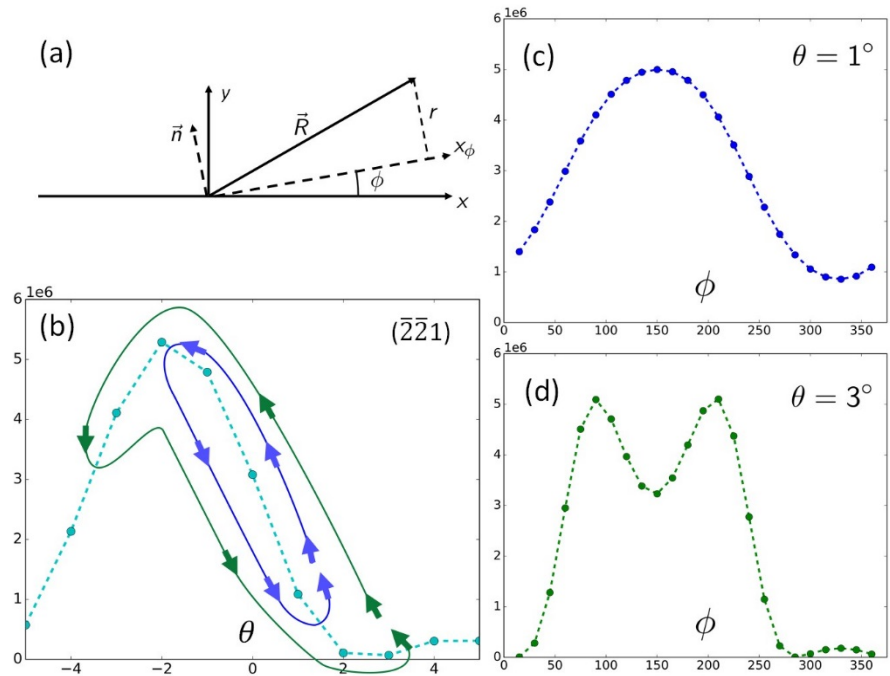

**Figure S4** Discussion of the effects of rotating the tilt-axis around z-axis: (a) vector $\vec{R}$ denoted the position of the diffraction spot in consideration, as the tilt-axis $x_\varphi$ rotated by the angle of φ, the corresponding tilt-arm length became $r = \vec{R} \cdot \vec{n}$, where $\vec{n}$ is the unit vector normal to tilt-axis, therefore, the effect of tilt-angle θ changed periodically with in-plane rotation-angle φ, (b) the relationship of simulated intensity of $(\bar{2}\bar{2}1)$ and tilt angle depicted variation of the corresponding reciprocal rel-rod in z direction, blue and green arrows indicated how the intersection, of Ewald sphere and the selected reciprocal rel-rod, moved with different ReSe$_2$ sample orientations in (c) and (d), which were acquired from simulations at fixed tilt-angle 1° and 3° with varying rotation-angle φ, ranging from 15° to 360°, at a step of 15°.

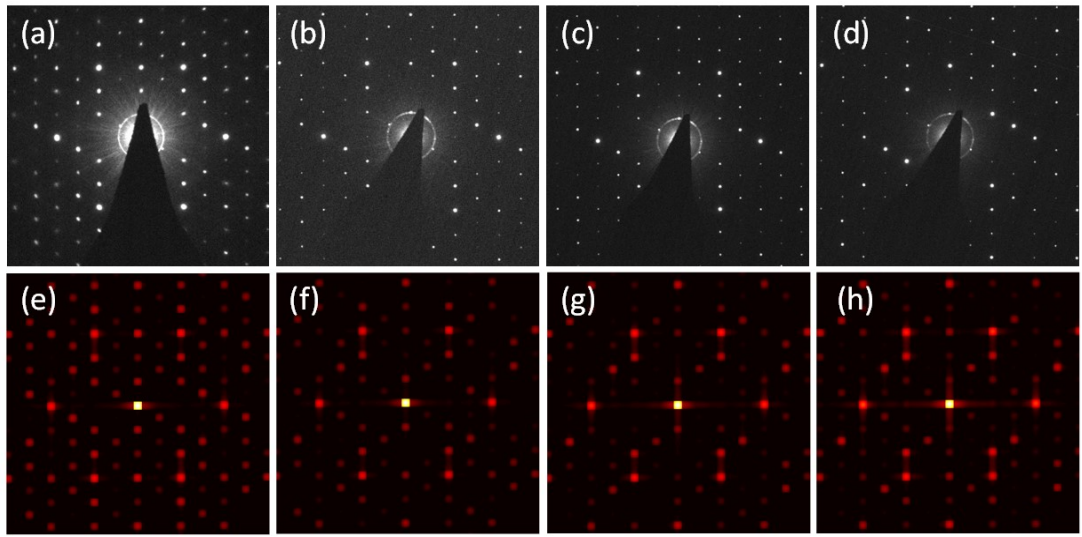

**Figure S5** (a)-(d) experimental diffraction patterns for 1-4 L ReSe$_2$, (e)-(h) simulated diffraction patterns for 1-4 L downside-oriented ReSe$_2$ without tilting. As we can see, as layer number increased, the experimental diffraction patterns and the simulated ones became less consistent, which could be attributed to that shape factors are more sensitive to tilt-angle with increasing layer number.

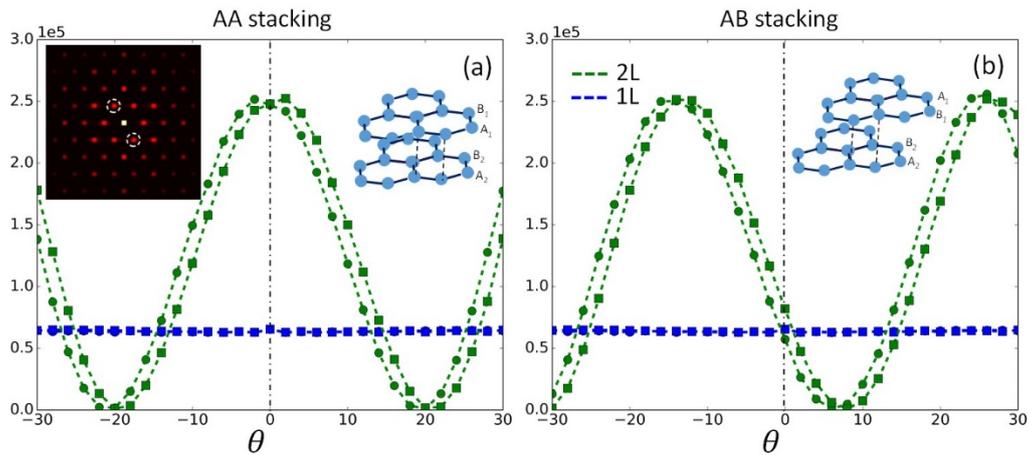

**Figure S6** Generalization of our method to high symmetry 2D materials, e.g., graphene. (a)&(b) show how the simulated intensities of a Friedel pair (indicated by a pair of white dashed-circle in the inset of (a) ) of mono- and bi- layer graphene vary with tilt-angle for AA stacking and AB stacking respectively. For both stacking configurations, there was a phase shift within the Friedel pair for bilayer graphene, furthermore, AB-stacking bilayer graphene was similar to triclinic crystal system, and there was a mismatch in intensity even without tilting, but in AA-stacking configuration, tilt-angle was required to distinguish monolayer from bilayer graphene by utilizing the corresponding intensity mismatch. Simulations were carried out under acceleration voltage of 60kV.